\documentclass[fleqn]{article}
\usepackage{espcrc2}

\begin{document}

\title{%
\bf Vector Manifestation in Hot Matter}
\author{{\bf Masayasu Harada}\address[SUNY]{%
Department of Physics and Astronomy,
SUNY at Stony Brook, Stony Brook, NY 11794, USA}%
\thanks{Present address: Department of Physics, Seoul National
University, Seoul, 151-742, Korea.}
and
{\bf Chihiro Sasaki}\address[Nagoya]{%
Department of Physics, Nagoya University,
Nagoya, 464-8602, Japan.}
}

\begin{abstract}
Based on the hidden local symmetry (HLS) Lagrangian as an effective
field theory of QCD, we find that the chiral symmetry restoration for
hot QCD can be realized through the Vector Manifestation where
the $\rho$ meson becomes
massless degenerate with $\pi$ as the chiral partner.
This is done by including, in addition to the hadronic thermal effects
due to the $\pi$- and $\rho$-loops, the intrinsic temperature
dependences of the parameters of the HLS Lagrangian through the
matching of the HLS with the underlying QCD.
\end{abstract}

\maketitle

\section{Introduction}

Vector meson mass in hot and/or dense matter is one of the most
interesting physical quantities in studying the hot and/or dense
QCD where the chiral symmetry is expected to be restored
(for reviews, see, e.g., 
Refs.~\cite{restoration,Brown-Rho:96,%
Brown-Rho:01b,Rapp-Wambach:00}).
The BNL Relativistic Heavy Ion Collider (RHIC) has started to measure
several effects in hot and/or dense matter.
Especially, the light vector meson mass is important for analysing the
dilepton spectra in RHIC.
In Refs.~\cite{Brown-Rho:91,Brown-Rho:96} 
it was proposed that the $\rho$-meson mass
scales like the pion decay constant and
vanishes at the chiral phase transition point
in hot and/or dense matter.

To study the $\rho$ mass in hot matter
it is useful to use models including the $\rho$ meson.
Among several such models we use
the model based on the hidden local 
symmetry (HLS)~\cite{BKUYY=BKY:88}
which successfully
reproduces the phenomena of 
$\rho$-$\pi$ system at zero temperature.
The HLS model is a natural extension of 
the nonlinear sigma model,
and reduces to it in the low-energy region.
It was shown~\cite{equivalence}
that the HLS model is equivalent to other models for vector mesons
at tree level.
We should stress here that,
as first pointed by Georgi~\cite{Georgi} and developed further
in Refs.~~\cite{HY,Tanabashi,HY:matching,HY:PR},
{\it thanks to the gauge symmetry in the HLS model, 
we can perform a systematic loop expansion including the vector mesons
in addition to the pseudoscalar
mesons}.

Several groups~\cite{Lee-Song-Yabu:95=Song-Koch:96,%
Harada-Shibata,Rapp-Wambach:00}
studied the 
$\rho$ mass in hot matter
using the HLS model.
Most of them included only the thermal effect of $\pi$ and
dropped that of $\rho$ itself.
In Ref.~\cite{Harada-Shibata},
the first application 
of the systematic chiral perturbation with
HLS~\cite{Georgi,HY,Tanabashi,HY:matching,HY:PR}
in hot matter
was made.
There hadronic thermal effects of $\rho$ and $\pi$ were included
at one loop
and the $\rho$ mass was shown to increase with temperature $T$
at low temperature.

In the analysis done in Ref.~\cite{Harada-Shibata}
the parameters of the Lagrangian at $T=0$ were used by
assuming no temperature dependences of them.
When we naively extrapolate the results in Ref.~\cite{Harada-Shibata}
to the critical temperature,
the resultant axialvector and vector current correlators do not agree
with each other.
Disagreement between these correlators
is obviously inconsistent with the chiral restoration in QCD.
However, the parameters of the HLS Lagrangian should be determined by
the underlying QCD.
As was shown in Ref.~\cite{HY:matching},
{\it the bare parameters of the (bare) HLS Lagrangian}
defined at the matching scale $\Lambda$ for $N_f = 3$ at $T=0$ 
{\it are determined by matching the HLS with the underlying QCD}
at $\Lambda$
through the Wilsonian matching conditions:
This was done by matching the current
correlators in the HLS with those derived by the operator product
expansion (OPE) in QCD.
Since the current correlators by the OPE at non-zero temperature
depend on the temperature
(see, e.g., Refs.~\cite{Adami-Brown,Hatsuda-Koike-Lee}),
the application of the Wilsonian matching 
to the hot matter calculation implies that
the bare parameters of the HLS 
do depend on the temperature,
which we call the 
{\it intrinsic temperature dependences}
in contrast to the hadronic thermal effects.
We stress here that
the above disagreement between the current correlators 
is cured by including
the intrinsic temperature dependences.

In Ref.~\cite{HY:VM},
on the other hand,
{\it the vector manifestation (VM)}
is proposed as a new pattern
of the Wigner realization of chiral
symmetry, in which 
the chiral symmetry is restored at the critical point by 
{\it the massless degenerate pion (and its flavor
partners) and the $\rho$ meson (and its flavor partners) as
the chiral partner},
in sharp contrast to the traditional manifestation \`a la linear
sigma model where the symmetry is restored by the degenerate pion and
the scalar meson.
It was shown that
VM actually takes place 
in the large $N_f$ QCD
through the Wilsonian matching.
Since the VM is a general property in the chiral restoration
when the HLS can be matched with the underlying QCD at the critical
point,
it was then 
suggested~\cite{HY:VM,Brown-Rho:01a,Brown-Rho:01b}
that the VM may be applied to the chiral restoration 
in hot and/or dense matter.

In this paper, we demonstrate that
{\it the VM can in fact occur in the chiral symmetry restoration 
in hot matter},
using the HLS as an effective field theory of QCD.
Here we determine 
the intrinsic temperature dependences of the
bare parameters of the HLS through the Wilsonian
matching in hot matter,
and convert them to the intrinsic temperature dependences of the
on-shell parameters by including the quantum effects through the
Wilsonian RGE's for the HLS parameters~\cite{HY:letter,HY:matching}.
Then, we separately include 
the hadronic thermal effects to obtain physical quantities
by explicitly calculating the $\pi$- and $\rho$- thermal loops.

\section{Hidden Local Symmetry}

Let us first describe the HLS model
based on the
$G_{\rm global} \times H_{\rm local}$ symmetry, where
$G = \mbox{SU($N_f$)}_{\rm L} \times 
\mbox{SU($N_f$)}_{\rm R}$  is the 
global chiral symmetry and 
$H = \mbox{SU($N_f$)}_{\rm V}$ is the HLS.
The basic quantities 
are the gauge boson 
$\rho_\mu$
and two 
variables 
\begin{eqnarray}
&&
\xi_{\rm L,R} = e^{i\sigma/F_\sigma} e^{\mp i\pi/F_\pi}
\ ,
\end{eqnarray}
where $\pi$
denotes the pseudoscalar Nambu-Goldstone (NG) boson
and $\sigma$~\footnote{
Note that this $\sigma$ is different with the scalar meson
in the linear sigma model.
}
the NG boson absorbed into $\rho_\mu$ 
(longitudinal $\rho$).
$F_\pi$ and $F_\sigma$ are relevant decay constants, and
the parameter $a$ is defined as
$a \equiv F_\sigma^2/F_\pi^2$.
The transformation property of $\xi_{\rm L,R}$ is given by
\begin{eqnarray}
&&
\xi_{\rm L,R}(x) \rightarrow \xi_{\rm L,R}^{\prime}(x) =
h(x) \xi_{\rm L,R}(x) g^{\dag}_{\rm L,R}
\ ,
\end{eqnarray}
where $h(x) \in H_{\rm local}$ and 
$g_{\rm L,R} \in G_{\rm global}$.
The covariant derivatives of $\xi_{\rm L,R}$ are defined by
\begin{eqnarray}
&&
D_\mu \xi_{\rm L} =
\partial_\mu \xi_{\rm L} - i g \rho_\mu \xi_{\rm L}
+ i \xi_{\rm L} {\cal L}_\mu
\ ,
\nonumber\\
&&
D_\mu \xi_{\rm R} =
\partial_\mu \xi_{\rm R} - i g \rho_\mu \xi_{\rm R}
+ i \xi_{\rm R} {\cal R}_\mu
\ ,
\label{covder}
\end{eqnarray}
where $g$ is the HLS gauge coupling, and
${\cal L}_\mu$ and ${\cal R}_\mu$ denote the external gauge fields
gauging the $G_{\rm global}$ symmetry.

The HLS Lagrangian is given by~\cite{BKUYY=BKY:88}
\begin{equation}
{\cal L} = F_\pi^2 \, \mbox{tr} 
\left[ \hat{\alpha}_{\perp\mu} \hat{\alpha}_{\perp}^\mu \right]
+ F_\sigma^2 \, \mbox{tr}
\left[ 
  \hat{\alpha}_{\parallel\mu} \hat{\alpha}_{\parallel}^\mu
\right]
+ {\cal L}_{\rm kin}(\rho_\mu) \ ,
\label{Lagrangian}
\end{equation}
where ${\cal L}_{\rm kin}(\rho_\mu)$ denotes the kinetic term of
$\rho_\mu$ 
and
\begin{eqnarray}
&&
\hat{\alpha}_{\perp,\parallel}^\mu =
( D_\mu \xi_{\rm R} \cdot \xi_{\rm R}^\dag \mp 
  D_\mu \xi_{\rm L} \cdot \xi_{\rm L}^\dag
) / (2i)
\ .
\end{eqnarray}
When the kinetic term ${\cal L}_{\rm kin}(\rho_\mu)$
is ignored in the low-energy region,
the second term of Eq.(\ref{Lagrangian}) vanishes
by integrating out $\rho_\mu$
and only the first term remains.
Then, 
the HLS model is reduced to the nonlinear sigma model based on $G/H$.

At zero temperature $T=0$,
it was shown~\cite{Georgi,Tanabashi} that, thanks to the gauge
symmetry in the HLS, we can perform the systematic loop expansion
including the vector meson.  Here the expansion parameter is a ratio
of the $\rho$ meson mass to the chiral symmetry breaking scale
$\Lambda_\chi$~\cite{Georgi}
in addition to the ratio of the momentum $p$ to
$\Lambda_\chi$ as used in the ordinary chiral perturbation theory.
By assigning ${\cal O}(p)$ to the HLS gauge coupling 
$g$~\cite{Georgi,Tanabashi}, 
the Lagrangian in Eq.~(\ref{Lagrangian}) is counted as 
${\cal O}(p^2)$, and one-loop quantum corrections obtained from
the Lagrangian are counted as ${\cal O}(p^4)$.

Due to quantum corrections,
three parameters $F_\pi$, $F_\sigma$ and $g$
are renormalized
at one-loop level,
and depend on the renormalization scale
$\mu$~\cite{HY,HY:letter,HY:matching}.
Furthermore, at non-zero temperature $T >0$,
these parameters 
have the intrinsic temperature dependences.
We write both dependences explicitly as
$F_\pi(\mu;T)$, $a(\mu;T)$ and $g(\mu;T)$~\footnote{%
The renormalization scale $\mu$ and the temperature
$T$ are independent of each other in the present approach.
}.

To avoid confusion,
we use $f_\pi$ for the physical decay constant of $\pi$,
and $F_\pi$ for the parameter of the Lagrangian.
Similarly, 
$M_\rho$ denotes the parameter of the Lagrangian and
$m_\rho$ the $\rho$ pole mass.
For calculating the hadronic thermal corrections it is convenient to
adopt the on-shell renormalization scheme at $T=0$
as in Ref.~\cite{Harada-Shibata}.
Below, we use the following abbreviated notations:
\begin{eqnarray}
\lefteqn{
  F_\pi = F_\pi(\mu=0;T) \ , 
}
\nonumber\\
\lefteqn{
  g = g\mbox{\boldmath$\bigl($} 
  \mu = M_\rho(T);T
  \mbox{\boldmath$\bigr)$}
\ ,
\quad
  a = a\mbox{\boldmath$\bigl($} 
  \mu = M_\rho(T);T
  \mbox{\boldmath$\bigr)$}
\ ,
}
\label{on-shell para T}
\end{eqnarray}
where $M_\rho$ is determined from the on-shell condition:
\begin{eqnarray}
&&
  \hspace*{-0.5cm}
  M_\rho^2 = M_\rho^2(T) =
  a\mbox{\boldmath$\bigl($} 
  \mu = M_\rho(T);T
  \mbox{\boldmath$\bigr)$}
\nonumber\\
&& \quad
  \hspace*{-0.5cm}
  \times
  g^2\mbox{\boldmath$\bigl($} 
  \mu = M_\rho(T);T
  \mbox{\boldmath$\bigr)$}
  F_\pi^2\mbox{\boldmath$\bigl($} 
  \mu = M_\rho(T);T
  \mbox{\boldmath$\bigr)$}
\ .
\end{eqnarray}
Then, the parameter $M_\rho$ in this paper is renormalized in 
such a way that it becomes the pole mass at $T=0$.

\section{Hadronic Thermal Corrections}

Here we summarize the hadronic thermal effects to the 
decay constant of $\pi$ and the $\rho$ mass shown in
Ref.~\cite{Harada-Shibata}
where the temperature $T$ is assigned to be of ${\cal O}(p)$ 
following Ref.~\cite{Gasser-Leutwyler:87}.

The decay constant of $\pi$ is defined
through the longitudinal component of the axialvector current
correlator
at the low energy limit~\cite{Bochkarev-Kapusta}.
The hadronic thermal corrections from $\pi$ and $\rho$ are 
summarized as~\cite{Harada-Shibata}
\begin{eqnarray}
\lefteqn{
f_\pi^2(T)=
F_\pi^{2} - \frac{N_f}{2\pi^2} 
\left[
  I_2  - a J_1^2
  +\frac{a}{3M_\rho^2} \left( I_{4}-J_{1}^{4}\right)
\right] 
\ ,
}
\nonumber\\
\label{eq: fpi}
\end{eqnarray}
where $I_n$ and $J_m^n$ ($n$, $m$\,: integer)
are defined as
\begin{eqnarray}
&&
  I_{n}
  \equiv
  \int_0^\infty d{\rm k} \frac{{\rm k}^{n-1}}{e^{{\rm k}/T}-1}
\ ,
\nonumber\\
&&
  J_m^n
  \equiv
  \int_0^\infty d{\rm k} \frac{1}{e^{\omega/T}-1}
  \frac{{\rm k}^n}{\omega^m} 
\ ,
\label{function 1}
\end{eqnarray}
with $\omega \equiv \sqrt{{\rm k}^2 + M_\rho^2}$.
When we consider the low temperature region $T \ll M_\rho$ 
in Eq.~(\ref{eq: fpi}), only the $I_2$ term remains:
\begin{eqnarray}
&&
f_\pi^2(T) \approx
F_\pi^2 - (N_f/ 2\pi^2) I_2
\nonumber\\
&& \qquad\quad
= F_\pi^2 - N_f T^2 / 12
\ ,
\label{fpi: ChPT}
\end{eqnarray}
which is consistent with the result in
Ref.~\cite{Gasser-Leutwyler:87}.

We estimate 
the critical temperature 
by naively extrapolating 
the above result 
to the higher temperature
without including the intrinsic temperature dependences.
The critical temperature 
for $N_f=3$ 
is approximated as
\begin{eqnarray}
&&
T_c^{\rm(had)} \approx \sqrt{ 12/N_f }\, f_\pi(T=0) 
\nonumber\\
&& \qquad\quad
= 2 f_\pi(0) 
\simeq 180\,\mbox{MeV}
\ .
\label{Tc had}
\end{eqnarray}

In Ref.~\cite{Harada-Shibata}
$m_\rho$
is defined
by the pole of the longitudinal $\rho$ propagator
at rest frame:
\begin{eqnarray}
&&
m_\rho^2(T) = M_\rho^2 - 
\mbox{Re}\,\Pi_V^L( p_0=M_\rho , \vec{p}=0;T)
\ ,
\end{eqnarray}
where $\mbox{Re}\,\Pi_V^L$ denotes the real part of the longitudinal 
component of the $\rho$ two-point function at one-loop level.
Inside the one-loop correction $\mbox{Re}\,\Pi_V^L$ we replaced
$m_\rho$ by $M_\rho$, since the difference is of higher order.
The resultant thermal corrections are
summarized as~\cite{Harada-Shibata}
\begin{eqnarray}
\lefteqn{
m_\rho^2(T)
  = M_\rho^2 
  -
  \frac{N_f g^2}{2\pi^2}
  \Biggl[
    \frac{a^2}{12} \bar{G}_2
    - \frac{5}{4} J_1^2
    - \frac{33}{16} M_\rho^2 \, \bar{F}_3^2
  \Biggr]
\ ,
}
\nonumber\\
&&
\label{mrho at T}
\end{eqnarray}
where $J_1^2$ is defined in Eq.~(\ref{function 1}),
and
$\bar{F}_3^n$ and $\bar{G}_n$ are defined as
\begin{eqnarray}
&&
  \bar{F}_3^n \equiv
  \int_0^\infty d{\rm k} {\cal P}
  \frac{1}{e^{\omega /T}-1}
  \frac{4{\rm k}^{n}}{\omega (4\omega^2 - M_\rho^2)} 
\ ,
\nonumber \\
&& 
  \bar{G}_n \equiv
  \int_0^\infty d{\rm k} {\cal P}
  \frac{{\rm k}^{n-1}}{e^{{\rm k}/T}-1}
  \frac{4{\rm k}^2}{4{\rm k}^2 - M_\rho^2}
\ ,
\label{function 2}
\end{eqnarray}
with ${\cal P}$ denoting the principal part.
{}From this expression
it was shown~\cite{Harada-Shibata} that there is no $T^2$ term
in the low temperature region consistently with the result in
Ref.~\cite{Dey-Eletsky-Ioffe}.

\section{Intrinsic Temperature Dependences}

Let us now include the intrinsic temperature dependences
of $F_\pi$, $a$ and $g$ 
(and $M_\rho^2 = a g^2 F_\pi^2$) appearing in Eqs.~(\ref{eq: fpi})
and (\ref{mrho at T}).
To do that, we first determine
the bare parameters defined at the matching scale $\Lambda$ 
by extending the Wilsonian matching~\cite{HY:matching},
which was originally proposed for $T=0$,
to non-zero temperature.
We should note that, for the validity of the expansion in
the HLS, the matching scale $\Lambda$ must be smaller than the chiral
symmetry breaking scale $\Lambda_\chi$.
We match the axialvector and vector current correlators in the HLS
with those derived in the OPE for QCD at non-zero temperature
(see, e.g., Refs.~\cite{Adami-Brown,Hatsuda-Koike-Lee}).
The correlators in the HLS 
around the matching scale $\Lambda$
are well described by the same forms as those at 
$T=0$~\cite{HY:matching} with the bare parameters 
having the intrinsic temperature dependences:
\begin{eqnarray}
\Pi_A^{\rm(HLS)}(Q^2) &=&
\frac{F_\pi^2(\Lambda;T)}{Q^2} - 2 z_2(\Lambda;T)
\ ,
\nonumber\\
\Pi_V^{\rm(HLS)}(Q^2) &=&
\frac{
  F_\sigma^2(\Lambda;T)
  \left[ 1 - 2 g^2(\Lambda;T) z_3(\Lambda;T) \right]
}{
  M_\rho^2(\Lambda;T) + Q^2
}
\nonumber\\
&&
- 2 z_1(\Lambda;T)
\ ,
\label{Pi A V HLS}
\end{eqnarray}
where 
$M_\rho^2(\Lambda;T) \equiv g^2(\Lambda;T) F_\sigma^2(\Lambda;T)$ is
the bare $\rho$ mass,
and $z_{1,2,3}(\Lambda;T)$ are the bare coefficient
parameters of the relevant
${\cal O}(p^4)$ terms~\cite{Tanabashi,HY:matching,HY:PR}.
Matching the above correlators with those by the OPE in the same
way as done for $T=0$~\cite{HY:matching},
we determine the bare parameters
including the intrinsic temperature dependences,
which are then converted into those
of the on-shell parameters through the Wilsonian 
RGE's~\cite{HY:letter,HY:matching}.
As a result,
the parameters appearing in the hadronic
thermal corrections have the intrinsic temperature
dependences.
In this way we include 
{\it both the intrinsic and hadronic thermal effects}
together into the physical quantities.

\section{Vector Manifestation}

Now, we study the chiral restoration in hot matter.
Here we assume that the chiral broken phase is in the 
confining phase, i.e., the critical temperature $T_c$
for chiral
phase transition is not larger than the critical temperature for
confinement-deconfinement phase transition,
and the
hadronic picture is valid.
When the symmetry is completely restored, the HLS is not applicable.
We approach to the critical temperature from the broken phase
where the HLS is applicable.
At the moment we assume that the expansion parameter
$M_\rho/\Lambda_\chi$ is small near $T_c$.
It turns out that it is actually small since $M_\rho\rightarrow0$
when $T\rightarrow T_c$, as we will show below.
We first consider the Wilsonian matching 
at the critical temperature $T_c$ for $N_f = 3$
with assuming
that $\langle \bar{q} q \rangle$ approaches to $0$ 
continuously for $T \rightarrow T_c$~\footnote{%
It is known that there is no Ginzburg-Landau type phase transition for
$N_f=3$ (see, e.g., Refs.~\cite{restoration,Brown-Rho:96}).
There may still be a possibility of non-Ginzburg-Landau type
continuous phase transition such as the conformal phase
transition~\cite{Miransky-Yamawaki}.
When the Wilsonian matching can be applicable for $N_f=2$,
the VM should occur.
}.
In such a case,
the axialvector and vector current correlators by the OPE
approach to each other, and agree at $T_c$.
Then through the Wilsonian matching we require that the 
correlators in Eq.~(\ref{Pi A V HLS}) agree with
each other.
As was shown in Ref.~\cite{HY:VM} for large $N_f$ chiral
restoration, 
this agreement is satisfied if the following conditions are 
met~\footnote{%
  We should note that 
  we can take $T \rightarrow T_c$ limit with $\Lambda$ fixed
  in Eq.~(\ref{g a z12:VMT}) since 
  $\Lambda$ and $T$ in the bare
  parameters of the HLS are independent of each other.
}:
\begin{eqnarray}
&&
g(\Lambda;T) \mathop{\longrightarrow}_{T \rightarrow T_c} 0 \ ,
\qquad
a(\Lambda;T) \mathop{\longrightarrow}_{T \rightarrow T_c} 1 \ ,
\nonumber\\
&&
z_1(\Lambda;T) - z_2(\Lambda;T) 
\mathop{\longrightarrow}_{T \rightarrow T_c} 0 \ .
\label{g a z12:VMT}
\end{eqnarray}

As we explained above,
the conditions for the bare parameters
$g(\Lambda;T_c) =0$ and $a(\Lambda;T_c) = 1$ are converted into the
conditions for the on-shell parameters through the Wilsonian RGE's.
Since $g=0$ and $a=1$ are separately the fixed points of the RGE's for
$g$ and $a$~\cite{HY:letter},
the on-shell parameters also satisfy
$(g,a)=(0,1)$, and thus $M_\rho = 0$.

Let us include the hadronic thermal effects to obtain the $\rho$ pole
mass.
Here we extrapolate the result in Eq.~(\ref{mrho at T}) 
to the higher temperature {\it with including the intrinsic
temperature dependences of the parameters}.
Noting that 
$\bar{G}_2 \rightarrow \pi^2 T^2/6$,
$J_1^2 \rightarrow \pi^2 T^2/6$ and
$M_\rho^2 \bar{F}_3^2 \rightarrow 0$
for $M_\rho \rightarrow 0$,
Eq.~(\ref{mrho at T}) for $M_\rho \ll T$ 
reduces to
\begin{eqnarray}
&& m_\rho^2(T)
  = M_\rho^2 +
  g^2 \frac{N_f }{2\pi^2} \frac{15 - a^2}{12} \frac{\pi^2}{6} T^2
\ .
\label{mrho at T 2}
\end{eqnarray}
Since $a \simeq 1$ near the restoration point,
the second term is positive. 
Then the $\rho$ pole mass $m_\rho$
is bigger than the parameter
$M_\rho$ due to the hadronic thermal corrections.
Nevertheless, 
{\it the intrinsic temperature dependence determined by the
Wilsonian matching requires
that the $\rho$ becomes massless at the
critical temperature}:
\begin{eqnarray}
&&
m_\rho^2(T)
\mathop{\longrightarrow}_{T \rightarrow T_c} 0 \ ,
\end{eqnarray}
since the first term in Eq.~(\ref{mrho at T 2})
vanishes as $M_\rho\rightarrow 0$, and the second
term also vanishes since $g\rightarrow 0$ for $T \rightarrow T_c$.
This implies that, as was suggested in 
Refs.~\cite{HY:VM,Brown-Rho:01a,Brown-Rho:01b},
{\it the vector manifestation (VM) actually
occurs at the critical
temperature}.
This is the main result of this paper, which is
consistent with the picture shown in
Refs.~\cite{Brown-Rho:91,Brown-Rho:96,Brown-Rho:01a,Brown-Rho:01b}.
We should stress here that the above $m_\rho(T)$ is the $\rho$
pole mass, which is important
for analyzing the dilepton spectra in RHIC experiment.
It is noted~\cite{HY:VM} that although conditions for 
$g(\Lambda;T)$ and $a(\Lambda;T)$ in Eq.~(\ref{g a z12:VMT})
coincide with the Georgi's vector
limit~\cite{Georgi}, the VM ($f_\pi \rightarrow 0$)
should be distinguished from Georgi's
vector realization~\cite{Georgi}.

\section{Critical Temperature}

Let us determine the critical temperature.
For $T > 0$ the thermal averages of the 
Lorentz non-scalar 
operators such as 
$\bar{q} \gamma_\mu D_\nu q$ exist
in the OPE~\cite{Hatsuda-Koike-Lee}.
Since these 
are smaller than the main term 
$1 + \alpha_s/\pi$,
we expect that they give only
small corrections to 
the value of $T_c$,
and neglect them here.
Then, the Wilsonian matching condition to determine the bare parameter 
$F_\pi(\Lambda;T_c)$ is obtained from that in Eq.~(4.5) of
Ref.~\cite{HY:matching} by taking $\langle \bar{q} q \rangle=0$ and
including a possible temperature dependence of the gluonic condensate:
\begin{eqnarray}
\lefteqn{
\frac{F_\pi^2(\Lambda;T_c)}{\Lambda^2} 
= \frac{1}{8\pi^2}
\left[
  1 + \frac{\alpha_s}{\pi}
  + \frac{2\pi^2}{3} 
    \frac{
      \left\langle 
        \frac{\alpha_s}{\pi} G_{\mu\nu} G^{\mu\nu}
      \right\rangle_{T_c}
    }{ \Lambda^4 }
\right]
\ .
}
\nonumber\\
\label{fp Tc WM}
\end{eqnarray}
The on-shell parameter $F_\pi(0;T_c)$ 
is determined 
through 
the
Wilsonian RGE~\cite{HY:letter,HY:matching} 
for $F_\pi$ with taking $(g,a)=(0,1)$.
As for large $N_f$~\cite{HY:letter,HY:VM},
the result is given by
\begin{eqnarray}
&&
\frac{F_\pi^2(0;T_c)}{\Lambda^2} = 
\frac{F_\pi^2\left(\Lambda;T_c\right)}{\Lambda^2}
- \frac{N_f}{2(4\pi)^2} \ .
\label{RGE for fpi2 at vector limit at Tc}
\end{eqnarray}
We need to include the hadronic thermal effects
to obtain the relation between the parameter $F_\pi(0;T_c)$ and 
the order parameter $f_\pi(T_c)$.
Here we extrapolate the hadronic thermal effect
shown in Eq.~(\ref{eq: fpi}) to higher temperature
{\it with including the intrinsic thermal effect}.
Then, taking $M_\rho \rightarrow 0$ and 
$a \rightarrow 1$ in Eq.~(\ref{eq: fpi}), we obtain
\begin{eqnarray}
&&
0 = f_\pi^2(T_c) =
F_\pi^2(0;T_c) - N_f T_c^2 /24
\ .
\label{Fp: WM}
\end{eqnarray}
Here we should note that the coefficient of $T^2$ in the second term
is a half of that in Eq.~(\ref{fpi: ChPT})
which is an approximate form for $T \ll M_\rho$ taken with assuming
that the $\rho$ does not become light.
On the other hand, here the factor $1/2$ appears from 
the contribution of $\sigma$ (longitudinal $\rho$)
which becomes the real NG boson at $T=T_c$ due to the VM
where the chiral restoration in QCD predicts 
$a \rightarrow 1$ and $g \rightarrow 0$ for
$T \rightarrow T_c$.
{}From Eq.~(\ref{Fp: WM}) together with Eqs.~(\ref{fp Tc WM})
and (\ref{RGE for fpi2 at vector limit at Tc}), $T_c$
is expressed as
\begin{eqnarray}
\lefteqn{
T_c = \sqrt{ \frac{24}{N_f} } F_\pi(0;T_c)
= 
\sqrt{ \frac{3 \Lambda^2}{N_f\pi^2} }
}
\nonumber\\
&&
\times
\left[
  1 + \frac{\alpha_s}{\pi}
  + \frac{2\pi^2}{3} 
    \frac{
      \left\langle 
        \frac{\alpha_s}{\pi} G_{\mu\nu} G^{\mu\nu}
      \right\rangle_{T_c}
    }{ \Lambda^4 }
  - \frac{N_f}{4}
\right]^{1/2}
\ .
\nonumber\\
\label{Tc VM}
\end{eqnarray}

We estimate the value of $T_c$ for $N_f=3$.
The value of the gluonic condensate near phase transition point
becomes about half of that at $T=0$~\cite{Miller,Brown-Rho:01b},
so we use 
$\left\langle \frac{\alpha_s}{\pi} G_{\mu\nu} G^{\mu\nu}
\right\rangle_{T_c} = 0.006 \,\mbox{GeV}^4$ obtained by multiplying
the value at $T=0$ shown in Ref.~\cite{SVZ} by $1/2$.
For the value of the QCD scale
$\Lambda_{\rm QCD}$ we use 
$\Lambda_{\rm QCD} = 400 \, \mbox{MeV}$~\footnote{%
This value of $\Lambda_{\rm QCD}$ is within the range of values
estimated in Ref.~\cite{Buras};
$\Lambda_{\overline{\rm MS}}^{(3)} = 297 \sim 457$\,MeV.
}
as a typical example.
For this value of $\Lambda_{\rm QCD}$, it was
shown~\cite{HY:matching} that the choice of 
$\Lambda = 1.1 \,\mbox{GeV}$ provides the predictions in good
agreement with experiment at $T=0$.
However, the matching scale may have the temperature dependence.
In the present analysis
we use 
$\Lambda=0.8$, $0.9$, $1.0$ and $1.1$\,GeV,
and determine $T_c$ from Eq.~(\ref{Tc VM}).
We show the resultant values in Table~\ref{tab:Tc}.
\begin{table}[htbp]
\begin{center}
\begin{tabular}{|c||r|r|r|r|}
\hline
 $\Lambda$ &  $0.8$ &  $0.9$ &  $1.0$ &  $1.1$ \\
\hline
 $T_c$     & $0.21$ & $0.22$ & $0.23$ & $0.25$ \\
\hline
\end{tabular}
\end{center}
\caption[]{\small
Estimated values of the critical temperature $T_c$ for
several choices of the value of the matching scale $\Lambda$.
Units of $\Lambda$ and $T_c$ are GeV.
}\label{tab:Tc}
\end{table}

We note that the estimated values of $T_c$ in Table~\ref{tab:Tc} are 
larger than that in Eq.~(\ref{Tc had})
which is obtained by naively
extrapolating the temperature dependence from the hadronic thermal
effects without including the intrinsic temperature dependences.
This is because the extra factor $1/2$ appears in the second term in
Eq.~(\ref{Fp: WM}) compared with that in Eq.~(\ref{Tc had}).
As we stressed below Eq.~(\ref{Fp: WM}), the factor $1/2$ comes from
the contribution of $\sigma$ (longitudinal $\rho$) which becomes
massless at the chiral restoration point.

\section{Summary and Discussions}

To conclude,
{\it by imposing the Wilsonian matching
of the HLS with the underlying QCD at the critical temperature,
where the chiral symmetry restoration takes place,
the vector manifestation (VM) necessarily occurs:
The vector meson mass becomes zero}.
Accordingly,
the light vector meson gives a large thermal correction to the pion
decay constant,
and the value of the critical temperature 
becomes larger than the value estimated by 
including only the $\pi$ thermal effect.
The result that the vector meson becomes light near the critical
temperature is consistent with the picture shown in 
Refs.~\cite{Brown-Rho:91,Brown-Rho:96,Brown-Rho:01a,Brown-Rho:01b}.

Several comments are in order:

As shown in Ref.~\cite{HY:VM},
in the VM only the
longitudinal $\rho$ couples to the vector current near the critical
point, 
and the transverse $\rho$ is decoupled from it.
The $A_1$ in the VM is resolved and/or decoupled from the axialvector
current near $T_c$ since there is no contribution in the vector
current correlator to be matched with the axialvector correlator.
We expect that the scalar meson is also resolved
and/or decoupled near $T_c$ since it in the VM is in the same
representation as the $A_1$ is in.
We also expect that
excited mesons 
are also resolved and/or decoupled.

The estimated values of $T_c$ shown in Table~\ref{tab:Tc}
as well as $T_c^{\rm(had)}$ in Eq.~(\ref{Tc had})
may be changed by higher order hadronic thermal 
effects,
as in the chiral perturbation analysis~\cite{Gerber-Leutwyler}.
On the other hand,
the VM at $T_c$ is governed by the fixed
point and not changed by higher
order effects.

The parameter $M_\rho^2$ 
in Eq.~(\ref{mrho at T})
presumably has an intrinsic temperature
dependence 
proportional to $T^2$ through the Wilsonian matching.
Since we studied the 
intrinsic 
dependences only at $T_c$,
we cannot definitely argue how $m_\rho(T)$ falls in $T$.
However, we think that 
$g^2(\Lambda;T)$ 
vanishes as $\langle \bar{q} q \rangle_T^2$
near $T_c$
in the VM. 
If $\langle \bar{q} q \rangle_T^2$ falls as $( 1 - T^2/T_c^2)$
near $T_c$,
then the $\rho$ pole mass $m_\rho^2(T)$ as well as the parameter 
$M_\rho^2(T)$ vanishes as $( 1 - T^2/T_c^2)$ which seems to agree 
with the behavior of $f_\pi^2(T)$.
In such a case the scaling property in the VM may be consistent 
with the Brown-Rho scaling 
$m_\rho(T)/m_\rho(0) \sim f_\pi(T)/f_\pi(0)$~\cite{Brown-Rho:91}.

Although we concentrated on the hot matter calculation in this paper,
the present approach can be applied to the general hot and/or dense
matter calculation.

At present,
there are no clear lattice data for the $\rho$ pole mass 
in hot matter.
Our result here will be checked by 
lattice analyses in future.~\cite{lattice}

In this paper we performed our
analysis at the chiral limit.
We need to include the explicit chiral symmetry breaking
effect from the current quark masses when we apply the present
analysis to the real QCD.  
In such a case, we need
the Wilsonian matching
conditions with including non-zero quark mass which have not yet
been established.
Here we expect that 
the qualitative structure obtained in the present analysis
will not be changed by the inclusion of
the current quark masses.

\section*{Acknowledgement}

We would like to thank Professor~Koichi Yamawaki for 
discussions and for critical reading of this manuscript.
We are also grateful to Professor~Mannque Rho for useful comments
and encouragements.
This work is supported in part by Grant-in-Aid for Scientific Research
(A)\#12740144 and
USDOE Grant \#DE-FG02-88ER40388.


\begin{thebibliography}{99}

\bibitem{restoration}
T.~Hatsuda and T.~Kunihiro,
%``QCD phenomenology based on a chiral effective Lagrangian,''
Phys.\ Rept.\  {\bf 247}, 221 (1994);
%[hep-ph/9401310].
%%CITATION = HEP-PH 9401310;%%
R.~D.~Pisarski,
%``Applications of chiral symmetry,''
hep-ph/9503330;
%%CITATION = HEP-PH 9503330;%%
T.~Hatsuda, H.~Shiomi and H.~Kuwabara,
%``Light Vector Mesons in Nuclear Matter,''
Prog.\ Theor.\ Phys.\  {\bf 95}, 1009 (1996);
%[nucl-th/9603043].
%%CITATION = NUCL-TH 9603043;%%
F.~Wilczek,
%``QCD in extreme conditions,''
%in {\it C99-06-27.1}
hep-ph/0003183.
%%CITATION = HEP-PH 0003183;%%

\bibitem{Brown-Rho:96}
G.E.~Brown and M.~Rho, 
%``Chiral restoration in hot and/or dense matter,''
Phys.\ Rept.\  {\bf 269}, 333 (1996).
%%CITATION = HEP-PH 9504250;%%

\bibitem{Brown-Rho:01b}
G.E.~Brown and M.~Rho, 
%``On the manifestation of chiral symmetry in nuclei and dense nuclear
%matter,''
hep-ph/0103102, to appear in Phys. Rept.
%%CITATION = HEP-PH 0103102;%%

\bibitem{Rapp-Wambach:00}
R.~Rapp and J.~Wambach,
%``Chiral symmetry restoration and dileptons in relativistic heavy-ion
%collisions,'' 
Adv.\ Nucl.\ Phys.\  {\bf 25}, 1 (2000).
%[hep-ph/9909229].
%%CITATION = HEP-PH 9909229;%%

\bibitem{Brown-Rho:91}
G.~E.~Brown and M.~Rho, 
%``Scaling effective Lagrangians in a dense medium,''
Phys.\ Rev.\ Lett.\  {\bf 66}, 2720 (1991).
%%CITATION = PRLTA,66,2720;%%

\bibitem{BKUYY=BKY:88}
M.~Bando, T.~Kugo, S.~Uehara, K.~Yamawaki and T.~Yanagida,
%``Is Rho Meson A Dynamical Gauge Boson Of Hidden Local Symmetry?,''
Phys.\ Rev.\ Lett.\  {\bf 54}, 1215 (1985);
%%CITATION = PRLTA,54,1215;%%
M.~Bando, T.~Kugo and K.~Yamawaki,
%``Nonlinear Realization And Hidden Local Symmetries,''
Phys.\ Rept.\  {\bf 164}, 217 (1988).
%%CITATION = PRPLC,164,217;%%

\bibitem{equivalence}
See, e.g.,
K.~Yamawaki,
%``Hidden Local Symmetry Versus Massive Yang-Mills In The Nonlinear
%Chiral Lagrangian,''
Phys.\ Rev.\ D {\bf 35}, 412 (1987);
%%CITATION = PHRVA,D35,412;%%
G.~Ecker, J.~Gasser, H.~Leutwyler, A.~Pich and E.~de Rafael,
%``Chiral Lagrangians For Massive Spin 1 Fields,''
Phys.\ Lett.\ B {\bf 223}, 425 (1989);
%%CITATION = PHLTA,B223,425;%%
M.~Tanabashi,
%``Formulations of spin 1 resonances in the chiral lagrangian,''
Phys.\ Lett.\ B {\bf 384}, 218 (1996);
%[hep-ph/9511367].
%%CITATION = HEP-PH 9511367;%%
M.~C.~Birse,
%``Effective chiral Lagrangians for spin-1 mesons,''
Z.\ Phys.\ A {\bf 355}, 231 (1996).
%[hep-ph/9603251].
%%CITATION = HEP-PH 9603251;%%

\bibitem{Georgi}
H.~Georgi,
%``New Realization Of Chiral Symmetry,''
Phys.\ Rev.\ Lett.\  {\bf 63}, 1917 (1989);
%%CITATION = PRLTA,63,1917;%%
%H.~Georgi,
%``Vector Realization Of Chiral Symmetry,''
Nucl.\ Phys.\ B {\bf 331}, 311 (1990).
%%CITATION = NUPHA,B331,311;%%

\bibitem{HY}
M.~Harada and K.~Yamawaki,
%``Hidden local symmetry at one loop,''
Phys.\ Lett.\ B {\bf 297}, 151 (1992).
%%CITATION = HEP-PH 9210208;%%

\bibitem{Tanabashi}
M.~Tanabashi,
%``Chiral perturbation to one loop including the rho meson,''
Phys.\ Lett.\ B {\bf 316}, 534 (1993).
%%CITATION = HEP-PH 9306237;%%

\bibitem{HY:matching}
M.~Harada and K.~Yamawaki,
%``Wilsonian matching of effective field theory with underlying QCD,''
Phys.\ Rev.\ D {\bf 64}, 014023 (2001).
%%CITATION = HEP-PH 0009163;%%

\bibitem{HY:PR}
M.~Harada and K.~Yamawaki,
to appear in Phys. Rept.

\bibitem{Lee-Song-Yabu:95=Song-Koch:96}
S.~H.~Lee, C.~Song and H.~Yabu,
%``Photon - vector meson coupling and vector meson properties at low
%temperature pion gas,''
Phys.\ Lett.\ B {\bf 341}, 407 (1995);
%[hep-ph/9408266].
%%CITATION = HEP-PH 9408266;%%
C.~Song and V.~Koch,
%``Pion electromagnetic form factor at finite temperature,''
Phys.\ Rev.\ C {\bf 54}, 3218 (1996).
%[nucl-th/9608010].
%%CITATION = NUCL-TH 9608010;%%

\bibitem{Harada-Shibata}
M.~Harada and A.~Shibata,
%``Pion decay constants and the rho meson mass at finite
%temperature  in the hidden local symmetry,'' 
Phys.\ Rev.\ D {\bf 55}, 6716 (1997).
%[hep-ph/9612358].
%%CITATION = HEP-PH 9612358;%%

\bibitem{Adami-Brown}
C.~Adami and G.~E.~Brown,
%``Order of the QCD transition and QCD sum rules,''
Phys.\ Rev.\ D {\bf 46}, 478 (1992);
%%CITATION = PHRVA,D46,478;%%
%C.~Adami and G.~E.~Brown,
%``Matter under extreme conditions,''
Phys.\ Rept.\  {\bf 234}, 1 (1993).
%%CITATION = PRPLC,234,1;%%

\bibitem{Hatsuda-Koike-Lee}
T.~Hatsuda, Y.~Koike and S.~Lee,
%``Finite temperature QCD sum rules reexamined: rho, omega and A1
%mesons,'' 
Nucl.\ Phys.\ B {\bf 394}, 221 (1993).
%%CITATION = NUPHA,B394,221;%%

\bibitem{HY:VM}
M.~Harada and K.~Yamawaki,
%``Vector manifestation of the chiral symmetry,''
Phys.\ Rev.\ Lett.\  {\bf 86}, 757 (2001).
%%CITATION = HEP-PH 0010207;%%

\bibitem{Brown-Rho:01a}
G.E.~Brown and M.~Rho, 
%``Hidden local symmetry, color flavor locking and BR scaling,''
nucl-th/0101015.
%%CITATION = NUCL-TH 0101015;%%

\bibitem{HY:letter}
M.~Harada and K.~Yamawaki,
%``Conformal phase transition and fate of the hidden local symmetry in
%large N(f) QCD,'' 
Phys.\ Rev.\ Lett.\  {\bf 83}, 3374 (1999).
%%CITATION = HEP-PH 9906445;%%

\bibitem{Gasser-Leutwyler:87}
J.~Gasser and H.~Leutwyler,
%``Light Quarks At Low Temperatures,''
Phys.\ Lett.\ B {\bf 184}, 83 (1987).
%%CITATION = PHLTA,B184,83;%%

\bibitem{Bochkarev-Kapusta} 
%This $f_\pi^2(T)$ is defined as in
A.~Bochkarev and J.~Kapusta,
%``Chiral symmetry at finite temperature: linear vs nonlinear
%$\sigma$-models,''
Phys.\ Rev.\ D {\bf 54}, 4066 (1996).
%[hep-ph/9602405].
%%CITATION = HEP-PH 9602405;%%

\bibitem{Dey-Eletsky-Ioffe}
M.~Dey, V.~L.~Eletsky and B.~L.~Ioffe,
%``Mixing Of Vector And Axial Mesons At Finite Temperature: An
%Indication Towards Chiral Symmetry Restoration,''
Phys.\ Lett.\ B {\bf 252}, 620 (1990).
%%CITATION = PHLTA,B252,620;%%

\bibitem{Miransky-Yamawaki}
V.~A.~Miransky and K.~Yamawaki,
%``Conformal phase transition in gauge theories,''
Phys.\ Rev.\ D {\bf 55}, 5051 (1997)
[Erratum-ibid.\ D {\bf 56}, 3768 (1997)].
%[hep-th/9611142].
%%CITATION = HEP-TH 9611142;%%

\bibitem{Miller}
D.~E.~Miller,
%``Gluon condensates at finite temperature,''
arXiv:hep-ph/0008031.
%%CITATION = HEP-PH 0008031;%%

\bibitem{SVZ}
M.A.~Shifman, A.I.~Vainstein and V.I.~Zakharov,
Nucl. Phys. B {\bf 147} 385 (1979);
Nucl. Phys. B {\bf 147} 448 (1979).

\bibitem{Buras}
A.~J.~Buras,
%``Weak Hamiltonian, CP violation and rare decays,''
hep-ph/9806471.
%%CITATION = HEP-PH 9806471;%%
%Ref.~\cite{Buras}; 

\bibitem{Gerber-Leutwyler}
P.~Gerber and H.~Leutwyler, Nucl. Phys. B {\bf 321}, 387 (1989).

\bibitem{lattice}
See, e.g., M.~Asakawa, T.~Hatsuda and Y.~Nakahara,
%``Maximum entropy analysis of the spectral functions in lattice QCD,''
Prog.\ Part.\ Nucl.\ Phys.\  {\bf 46}, 459 (2001);
%[arXiv:hep-lat/0011040].
%%CITATION = HEP-LAT 0011040;%%
F.~Karsch,
%``Lattice results on QCD thermodynamics,''
%arXiv:
hep-ph/0103314;
%%CITATION = HEP-PH 0103314;%%
and references cited therein.

\end{thebibliography}
\end{document}